\documentclass[conference,10pt]{IEEEtran}

\usepackage{amsmath}

\usepackage{amssymb,amsfonts}
\usepackage{algorithmic}
\usepackage{graphicx}
\usepackage{textcomp}
\usepackage{xcolor}
\def\BibTeX{{\rm B\kern-.05em{\sc i\kern-.025em b}\kern-.08em
    T\kern-.1667em\lower.7ex\hbox{E}\kern-.125emX}}
\usepackage[utf8]{inputenc}
\usepackage{booktabs}
\usepackage{graphicx}
\usepackage{subfigure}
\usepackage{units}
\usepackage{blindtext}
\usepackage{multirow}
\usepackage{xcolor,xspace}
\usepackage{amsmath}
\usepackage{paralist}
\usepackage{mdframed}
\usepackage{pict2e}
\usepackage{mdframed}
\usepackage{listings}
\usepackage{color}
\usepackage{threeparttable}
\usepackage{array}
\usepackage{booktabs}
\usepackage{pifont}
\usepackage{balance}
\usepackage{url}
\usepackage{tabularx}
\usepackage{caption}
\usepackage{lipsum}
\expandafter\def\csname ver@l3regex.sty\endcsname{}

\usepackage[ruled]{algorithm2e}
\usepackage{algorithmic}
\LinesNumbered

\usepackage{url}

\usepackage[
    n,
    advantage,
    operators,
    sets,
    adversary,
    landau,
    probability,
    notions,
    logic,
    ff,
    mm,
    primitives,
    events,
    complexity,
    asymptotics,
    keys]{cryptocode}

\usetikzlibrary{arrows, patterns, automata}

\newcommand{\casu}{{{\sf CASU}}\xspace}

\newcommand{\casutextunderlined}{{{\underline{C}ompromise \underline{A}voidance via \underline{S}ecure \underline{U}pdate}}\xspace}

\newcommand{\casusw}{{{\sf \casu{\it-SW}}}\xspace}
\newcommand{\casuhw}{{{\sf \casu{\it-HW}}}\xspace}

\newcommand{\vrased}{{{\sf\it VRASED}}\xspace}
\newcommand{\rata}{{{\sf\it RATA}}\xspace}

\newcommand{\pure}{{{\sf\it PURE}}\xspace}
\newcommand{\apex}{{{\sf\it APEX}}\xspace}

\newcommand{\ignore}[1]{}

\newcommand{\prv}{{\ensuremath{\sf{\mathcal Prv}}}\xspace}
\newcommand{\vrf}{{\ensuremath{\sf{\mathcal Vrf}}}\xspace}
\newcommand{\RA}{{\ensuremath{\sf{\mathcal RA}}}\xspace}

\newcommand{\chal}{{\ensuremath{\sf{\mathcal Chal}}}\xspace}

\renewcommand{\key}{\ensuremath{\mathcal K}\xspace}

\renewcommand\adv{\ensuremath{\sf{\mathcal Adv}}\xspace}

\newcommand{\update}{\ensuremath{\mathsf{\bf Update}^{\vrf}}\xspace}
\newcommand{\authenticate}{\ensuremath{\mathsf{\bf Auth}^{\prv}}\xspace}
\newcommand{\install}{\ensuremath{\mathsf{\bf Install}^{\prv}}\xspace}
\renewcommand{\verify}{\ensuremath{\mathsf{\bf Verify}^{\vrf}}\xspace}
\newcommand{\fw}{{\ensuremath{\sf{\mathcal S_{new}}}}\xspace}
\newcommand{\ow}{{\ensuremath{\sf{\mathcal S_{old}}}}\xspace}
\newcommand{\atoken}{{\ensuremath{\mathsf{ATok}}}\xspace}
\newcommand{\aack}{{\ensuremath{\mathsf{AAck}}}\xspace}

\newcommand{\download}{\ensuremath{\mathsf{\bf download}}\xspace}
\newcommand{\acknowledge}{\ensuremath{\mathsf{\bf acknowledge}}\xspace}

\mathchardef\mhyphen="2D

\newcommand{\toctou}{{\small TOCTOU}\xspace}

\newcommand{\hmac}{HMAC\xspace}
\newcommand{\ER}{\ensuremath{ER}\xspace}
\newcommand{\EP}{\ensuremath{EP}\xspace}
\newcommand{\bEP}{\ensuremath{bEP}\xspace}
\newcommand{\TCR}{\ensuremath{TCR}\xspace}
\newcommand{\SR}{\ensuremath{SF}\xspace}
\newcommand{\IVTR}{\ensuremath{IVTR}\xspace}
\newcommand{\IVT}{\ensuremath{IVT}\xspace}
\newcommand{\daddr}{\ensuremath{D_{addr}}\xspace}
\newcommand{\wen}{\ensuremath{W_{en}}\xspace}
\newcommand{\pc}{\ensuremath{PC}\xspace}
\newcommand{\dmaaddr}{\ensuremath{DMA_{addr}}\xspace}
\newcommand{\dmaen}{\ensuremath{DMA_{en}}\xspace}
\newcommand{\reset}{\ensuremath{reset}\xspace}
\newcommand{\ermin}{\ensuremath{\ER_{min}}\xspace}
\newcommand{\ermax}{\ensuremath{\ER_{max}}\xspace}
\newcommand{\ATR}{\ensuremath{ATR}\xspace}
\newcommand{\status}{\ensuremath{Status}\xspace}

\newcommand{\modmem}{\ensuremath{Mod\_Mem}\xspace}

\newtheorem{construction}{Construction}

\begin{document}


\title{\casu: \casutextunderlined \\ for Low-end Embedded Systems}


\author{\IEEEauthorblockN{Ivan De Oliveira Nunes}
\IEEEauthorblockA{\textit{Rochester Institute of Technology}}

\and
\IEEEauthorblockN{Sashidhar Jakkamsetti}
\IEEEauthorblockA{\textit{UC Irvine}}

\and
\IEEEauthorblockN{Youngil Kim}
\IEEEauthorblockA{\textit{UC Irvine}}

\and
\IEEEauthorblockN{Gene Tsudik}
\IEEEauthorblockA{\textit{UC Irvine}}
}





\maketitle

\begin{abstract}
Guaranteeing runtime integrity of embedded system software is an open problem. 
Trade-offs between security and other priorities (e.g., cost or performance) are inherent, and resolving them is both challenging and important. 
The proliferation of runtime attacks that introduce malicious code (e.g., by injection) into embedded devices has prompted a range of mitigation techniques. 
One popular approach is Remote Attestation (\RA), whereby a trusted entity (verifier) checks the current software state of an untrusted remote device (prover). 
RA yields a timely authenticated snapshot of prover state that verifier uses to decide whether an attack occurred.

Current RA schemes require verifier to explicitly initiate \RA, based on some unclear criteria. 
Thus, in case of prover's compromise, verifier only learns about it late, upon the next \RA instance. 
While sufficient for compromise detection, some applications would benefit from a more proactive, prevention-based approach. 
To this end, we construct \casu: Compromise Avoidance via Secure Updates.
\casu is an inexpensive hardware/software co-design enforcing: (i) runtime software immutability, thus precluding any illegal software modification, and (ii) authenticated updates as the sole means of modifying software.
In CASU, a successful \RA instance serves as a proof of successful update, and continuous subsequent software integrity is implicit, due to the runtime immutability guarantee. 
This obviates the need for \RA in between software updates and leads to unobtrusive integrity assurance with guarantees akin to those of prior \RA techniques, with better overall performance.
\end{abstract}

\section{Introduction}\label{sec:intro}
Over the past two decades, Internet-of-Things (IoT) devices and Cyber-Physical Systems (CPS) have become very popular. 
They are deployed in many everyday settings, including both private (e.g., homes, offices, and factories) and public (e.g., cultural, entertainment, and transportation) spaces. 
They are also widely used in farming, industrial, and vehicular automation. 
These devices often collect sensitive information and perform safety-critical tasks. 
Also, in many cases, they are both interconnected and connected to the global Internet.
They are usually implemented atop low-end microcontroller units (MCUs) that have very stringent cost, size, and energy constraints, and unlike their higher-end counterparts, have no (or few) security features.
It is thus not at all surprising that these embedded devices (sensors, actuators, and hybrids) have become attractive attack targets.

In particular, code injection attacks \cite{FC08,runtime_attacks_sok,cowan2000buffer,owasp_attacks} represent a real and prominent threat to low-end devices. 
Embedded systems software is mostly written in C, C++, or Assembly -- languages that are very prone to errors. 
Code injection attacks exploit these errors to cause buffer overflows and inject malicious code into the existing software or somewhere else in the device memory.

Some previous results considered such attacks in low-end devices and proposed security techniques such as Remote Attestation (\RA) \cite{smart,sancus,vrasedp,simple,tytan,trustlite}, as well as proofs of remote software updates and memory erasure~\cite{pure,verify_and_revive,asokan2018assured}. 
\RA  aims to detect compromise by authenticated measurement of the device's current software state. 
However, it has considerable runtime costs since it requires computing a cryptographic function (usually, a Message Authentication Code (MAC)) over the entire software. 
A recent result, \rata~\cite{rata}, minimized the cost of \RA by measuring a constant-size memory region that reflects the time of last software modification (legal or otherwise).
\rata achieved that by introducing a hardware security monitor that securely logs each modification time to that region. 

Regardless of their specifics, \RA techniques only \underline{detect} code modifications {\bf after the fact}. 
They cannot prevent them from taking place. 
Hence, there could be a sizeable window of time between the initial compromise and the next \RA instance when the compromise would be detected.

To this end, the goal of this paper is to take a more proactive, prevention-based approach to avoid potential compromise.
It constructs \casu: \casutextunderlined, which consists of two main components.
First is a simple hardware security monitor that is formally verified. 
It performs two functions: (1) blocks all modifications to the specific program memory region where the software resides, and (2) prevents anything stored outside that region from executing. 
It runs independently from (in parallel with) the MCU core, without modifying the latter. 
This thwarts all code injection attacks. 
However, it is unrealistic to prohibit all modifications to program memory, since genuine software updates need to be installed during the device's lifetime.  
Otherwise, the software could be housed in ROM or the entire device would function as an ASIC (Application Specific Integrated Circuit). 
Therefore, \casu second component is a secure remote software update scheme. 

The key benefit of \casu is maintaining constant software integrity without repeated \RA measurements while allowing genuine secure software updates. 
Specifically, it guarantees that, between any two successive secure updates, device software is immutable. 
However, the device liveness can be ascertained at any time by repeating the latest update, which essentially represents \RA. 

\noindent~The intended contributions of \casu are:
\begin{compactenum}  
    \item A tiny formally verified hardware monitor that guarantees benign (authorized) software immutability and prevents the execution of any unauthorized code.
    \item A scheme to enable secure software updates when authorized by a trusted 3rd party.
    \item An open-source \casu prototype built atop a commodity low-end MCU to demonstrate its low cost and practicality.
\end{compactenum}

\section{Preliminaries}\label{sec:background}
\subsection{Targeted Devices}\label{subsec:background_scope}
\begin{figure}[h]
    \centering
    \captionsetup{justification=centering}
    \includegraphics[width=0.8\columnwidth]{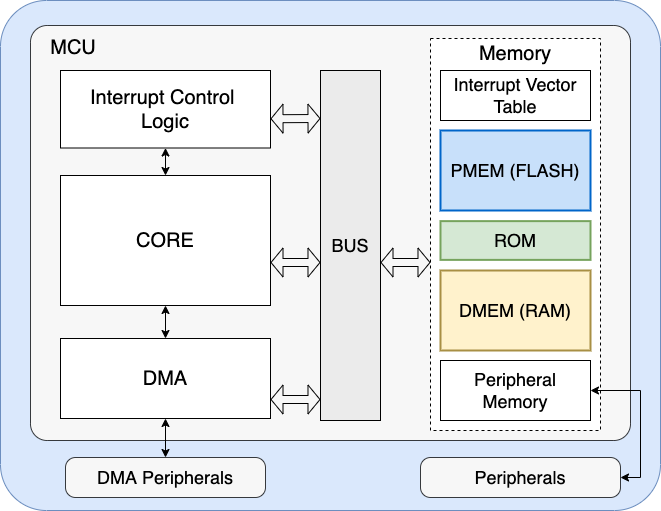}
    \caption{System architecture of a typical low-end MCU.}
    \label{fig:mcu}
\end{figure}

This paper focuses on CPS/IoT sensors and actuators (or hybrids thereof) with low computing power.
These are some of the smallest and weakest devices based on ultra-low-power single-core MCUs with only a few KBytes of memory. 
Two prominent examples are Atmel AVR ATmega \cite{atmel_specs} and TI MSP430\cite{TI-MSP430}, with $8$- and $16$-bit CPUs respectively, typically running at $1$-$16$MHz clock frequencies, with $\approx64$ KBytes of addressable memory.
Figure \ref{fig:mcu} shows a typical architecture of such an MCU. 
It includes a CPU core, a Direct Memory Access (DMA) controller, and an interrupt control logic connected to the memory via a bus. 
DMA is a hardware controller that can read/write to memory in parallel with the core. 
Main memory contains several regions: Interrupt Vector Table (IVT), program memory (PMEM), read-only memory (ROM), data memory (DMEM or RAM), and peripheral memory. 
IVT stores pointers to the Interrupt Service Routines (ISRs), where the execution jumps when an interrupt occurs; it also contains the Reset Vector pointer from where the core starts to execute, after a reboot. 
Application software is installed in PMEM and it uses DMEM for its stack and heap. 
ROM contains the bootloader and/or any immutable software hard-coded at manufacturing time. 

MCUs usually run software atop ``bare metal'' and execute instructions in place, i.e., directly from PMEM. 
They have neither memory management units (MMUs) to support virtualization, nor memory protection units (MPUs) for isolating memory regions. 
Therefore, privilege levels and isolation regimes used in higher-end devices and generic trusted execution environments (e.g., ARM TrustZone \cite{trustzone} or Intel SGX \cite{sgx}) are not viable. 

\noindent {\em NOTE:} Our initial implementation of \casu uses MSP430 MCU, a common platform for low-end embedded devices. 
One important factor in this choice is the public availability of an open-source MSP430 MCU design -- OpenMSP430~\cite{OpenMSP430}. 
Nonetheless, \casu is readily applicable to other low-end MCUs of the same class.

\subsection{Remote Attestation \& VRASED}\label{subsec:background_vrased}
\RA, mentioned above, allows a trusted entity (verifier = \vrf) to remotely measure current memory contents (e.g., software) of an untrusted embedded device (prover = \prv). 
\RA is usually realized as a simple challenge-response protocol:
\begin{compactenum}
 \item \vrf sends an \RA request with a challenge (\chal) to \prv.  
 \item \prv receives the request and computes an authenticate integrity check over its software memory region and \chal. The memory region can be either pre-defined or explicitly specified in the \RA request.
 \item \prv returns the result to \vrf.
 \item \vrf verifies the result and decides if \prv is in a valid state.
\end{compactenum}
Although several \RA techniques for low-end devices have been proposed, only very few offer any concrete (provable) security guarantees. 
The latter include SIMPLE\cite{simple}, \vrased~\cite{vrasedp}, and a variant of SANCUS\cite{sancus}. 
While SIMPLE, as its name suggests, is simple, it is a purely software-based \RA technique (meaning that no hardware modifications are needed) that only protects against remote attacks and does not support DMA.
Whereas, SANCUS is a purely hardware-based \RA technique which, though very fast, incurs a significant hardware cost over the baseline MCU.

\vrased~\cite{vrasedp} is a formally verified hybrid (hardware/software) \RA design comprising verified hardware and software sub-modules.  
The software sub-module, which is immutable (stored in ROM), implements the authenticated integrity function computed over some ``Attested Region'' (AR) of \prv memory (usually in PMEM).
Meanwhile, its hardware component assures that its software counterpart executes securely and that no function of the \RA secret key (\key) is ever leaked.
The authenticated integrity function is realized with a formally verified HMAC implementation from the HACL* cryptographic library \cite{hacl} used to compute: 

\vspace{-0.25cm}
{
\footnotesize
\begin{equation} \label{eq:vrased}
H = HMAC(KDF(\key, \chal), AR)
\end{equation}
}
\vspace{-0.5cm}

where $KDF(\key, \chal)$ is a one-time key derived from the received \chal and \key using a key derivation function.

\noindent {\em NOTE:} \casu uses \vrased to verify the update request before it installs the new software on the device. 
Specifically, \casu invokes \vrased to compute equation \ref{eq:vrased} on the new software and checks whether $H$ matches an authentication token sent in the update request. 
Consequently, \casu update verification inherits the security properties of \vrased.

\subsection{\toctou Attacks \& \toctou-Security}\label{subsec:background_vrasedta}
All \RA techniques share a common limitation: they yield no information about the state of \prv software during the time \textbf{between} two consecutive \RA instances. 
Consequently, it is impossible to detect the past presence of transient malware that: (1) infected \prv, (2) remained active for a while, and (3) at some later time erased itself and restored \prv software to its ``good'' state. 
This holds as long as (1)-(3) occur between two successive \RA instances.
This attack type is referred to as \emph{Time-Of-Check Time-Of-Use} (\toctou).

One recent technique, \rata~\cite{rata}, mitigates \toctou attacks with a minimal additional hardware component that securely logs the time of the last PMEM modification to a protected memory region called Latest Modification Time (LMT) that can not be modified by any software. 
LMT is then covered by the \RA function. 
Therefore, an \RA response captures both the current software state of \prv and the time of change to that state. 
Furthermore, \rata minimizes the computational cost of \RA for \prv, since, instead of attesting its entire software, it suffices for \prv to attest just the LMT.
This way, instead of computing a MAC over the entire PMEM, \prv computes it over a fixed-size (32-byte) $LMT$ region.

\noindent{\em NOTE:} In this paper, unlike \rata, \casu actively {\bf prevents} any modification to PMEM at runtime, unless it is a securely and causally authorized (by the trusted \vrf) software update.

\section{\casu Scheme \& Assumptions}
\subsection{Basics}
Similar to the typical \RA setting, \casu involves a low-end MCU (\prv) and verifier (\vrf).
The latter is a trusted higher-end device, e.g, a laptop, a smartphone, a smart home gateway, or a device manufacturer's back-end server. 
\vrf is responsible for initiating each software update request, verifying whether the update was successful, and keeping track of the latest successfully confirmed software update.
We assume a single \vrf for a given \prv. 
Also, \prv and \vrf are assumed to share a master secret key (\key) installed on \prv at manufacturing time. 
Our discussion focuses on the symmetric key setting, which is more practical for low-end MCUs.
Nonetheless, the use of public-key cryptography is possible with some cosmetic changes to \casu, provided that \prv has sufficient computing capabilities\footnote{In case of MSP430, based on our experimental attempts, neither generating nor even verifying public key signatures is viable.}.

\subsection{Secure Update Overview} \label{subsec:su_scheme}
At the time of its initial deployment, \vrf is assumed to know the software state (\ow) of \prv. 
When \vrf later wishes to update this software, it issues an update request, denoted by \update, to \prv. 
This request carries the new software \fw and a fresh authentication token \atoken, based on \fw. 

When \prv receives an \update, \ow invokes \casu, which handles the update process in two steps: (1) \authenticate verifies that \atoken is a fresh and timely token that corresponds to \fw, and (2) if the first step succeeds, \install replaces \ow with \fw and generates an authenticated acknowledgment (\aack). 
At this point, \casu terminates and control is given to \fw which must send \aack to \vrf.

\begin{figure}
	\centering
    \captionsetup{justification=centering}
    \includegraphics[width=0.85\columnwidth]{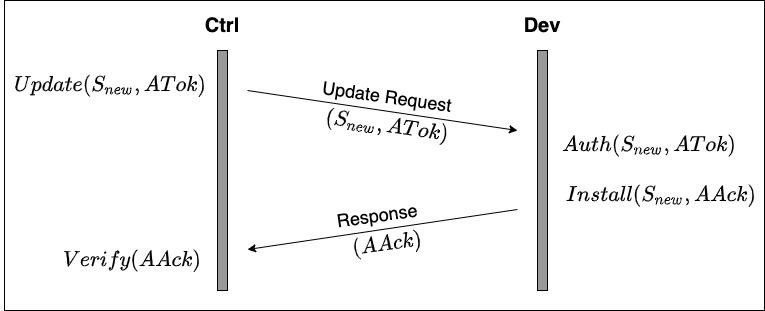}
    \caption{\casu Secure Update Protocol.}
    \label{fig:prot}
\end{figure}

Upon receiving \aack, \vrf executes the \verify procedure to check whether the \aack is a valid confirmation for the outstanding \update. 
If no \aack is received, or if \aack verification fails, \vrf assumes a failed update. 
Figure \ref{fig:prot} illustrates the interaction between \vrf and \prv. 
Protocol details are described in Section~\ref{sec:design} below.

\subsection{Adversary Model} \label{subsec:adv_model}
We consider an adversary, \adv, that controls the entire memory state of \prv, including PMEM (flash) and DMEM (RAM). 
It can attempt to write, read or execute any memory location. 
It can also attempt to remotely launch code injection attacks to modify \prv software. 
It may also divert the execution control-flow to ignore update requests, as well as attempt to extract any \prv secrets or forge update confirmations.

Furthermore, \adv can configure DMA controllers on \prv to read/write to any part of the memory while bypassing the CPU.
It can induce interrupts in an attempt to pause the update procedure, modify any part of the old or new software versions, or cause inconsistencies or race conditions. 
It might also eavesdrop on, and interfere, with network traffic between \vrf and \prv, in a typical Dolev-Yao manner~\cite{DolevYao}.

As common in most related work, physical attacks requiring adversarial presence are considered out of scope. 
This includes both non-invasive and invasive physical attacks. 
The former describes attacks whereby \adv physically reprograms \prv software using direct/wired interfaces, such as USB/UART, SPI, or I2C. 
The latter refers to inducing hardware faults, modifying code in ROM, extracting secrets via physical side-channels, and tampering with hardware. 
Protection against non-invasive attacks can be obtained via well-known features, such as a secure boot.
Whereas, protection against invasive attacks can be obtained via standard tamper-resistant techniques~\cite{ravi2004tamper}.

\section{\casu Design}\label{sec:design}
One of \casu main features is the prevention of all unauthorized software modifications to \prv software.
As mentioned earlier, the former can be trivially achieved by making all \prv software read-only, or by making \prv an ASIC.
However, this precludes all benign (authorized) updates. 
Therefore, it is essential to have a secure update mechanism.
The term ``authorized" refers to software installed on \prv physically at manufacture or deployment time, as well as each subsequent version installed via update request by \vrf.

From \vrf perspective, \casu guarantees that, once installed, authorized software on \prv remains unchanged until the 
next \vrf-initiated successful secure update. This is achieved via three features: 
\begin{compactenum}
    \item \textit{Authorized Software Immutability:} Except via a secure update (implemented within \casu trusted code), 
    authorized software cannot be modified. 
    \item \textit{Unauthorized Software Execution Prevention:} Only the memory containing the (immutable) authorized software is executable.
    \item \textit{Secure Update:} \vrf is the only entity that can authenticate to \prv to install new software. After an update, the previous version of the installed software is no longer authorized. 
\end{compactenum}
The first two features are realized by a hardware module, \casuhw, that runs in parallel with the CPU. 
It monitors a few CPU hardware signals and triggers an MCU reset if any violation is detected. 
The third feature is realized by a trusted code base (TCB), \casusw, that extends \vrased to authenticate incoming update requests containing new software to be installed (\fw) and an authorization token (\atoken) that must be issued by \vrf using the key \key pre-shared with \casu module within \prv.
If \atoken matches \fw, then \casusw installs \fw on \prv and produces an authenticated \aack, attesting to \vrf that a successful update occurred on \prv.

\begin{figure}
	\centering
    \captionsetup{justification=centering}
    \includegraphics[width=0.85\columnwidth]{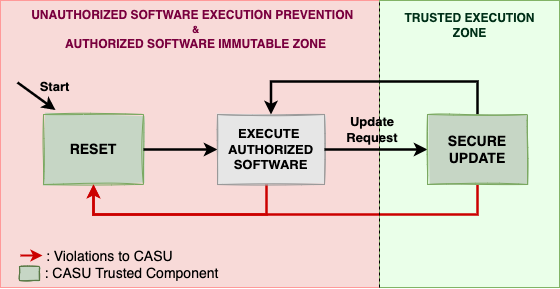}
    \caption{\casu Software Execution Flow.}
    \label{fig:overview}
\end{figure}

Figure \ref{fig:overview} depicts \casu software execution flow. 
After each boot or reset, it executes authorized software that was previously installed (either physically or via \casu Secure Update). 
In this state, \casuhw ensures software immutability and execution prevention of anything else.
However, when an update request is received, \casusw must be invoked to securely apply the update and re-configure \casuhw to protect the memory region where \fw is installed. 
Note that the update cannot be performed without invoking \casusw due to the immutability guarantee.

\begin{figure}
	\centering
    \captionsetup{justification=centering}
    \includegraphics[width=0.85\columnwidth]{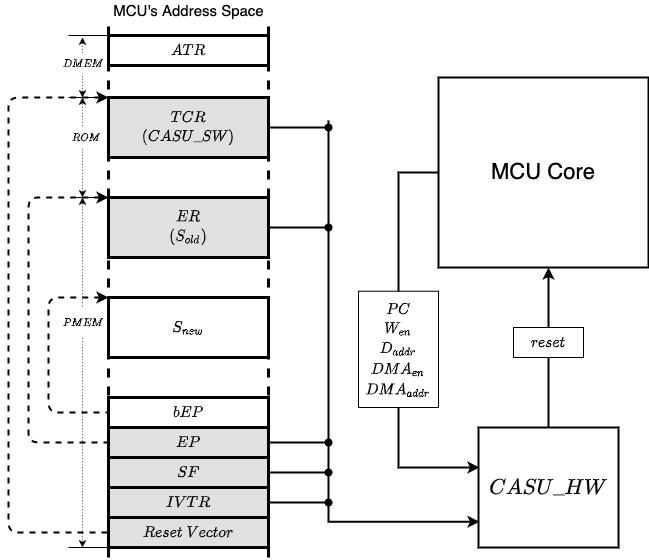}
    \caption{\casu System Architecture.}
    \label{fig:arch}
\end{figure}

\begin{table}
    \caption{Notation Summary} \label{table:notations}
    \footnotesize
    \linespread{0.8}
    \renewcommand{\arraystretch}{1.1}
    \begin{tabularx}{\columnwidth}{lX}
    \toprule
    Notation & Description    \\ 
    \midrule
    \pc	& 	Program Counter, points to the current instruction being executed \\
    \wen 	&	1-bit signal that indicates if MCU core is writing to memory \\
    \daddr 	&	Memory address where the MCU core is currently accessing  \\
    \dmaen &	1-bit signal that indicates if DMA is active \\
    \dmaaddr &	Memory address being accessed by DMA, when active     \\
    \reset	&	Signal that reboots the MCU when set to logic `1'\\
    \TCR	&	Trusted Code Region, a fixed ROM region storing \casusw \\
    \ER	&	Executable Region, a configurable memory region where authorized software is stored; $\ER = [\ermin, \ermax]$, where \ermin and \ermax are the boundaries of $ER$\\
    \EP	&	Executable Pointer, a fixed memory region storing current values of \ermin and \ermax\\
    \bEP	&	Buffer Executable Pointer, a fixed memory location used to save the boundaries of the memory region storing new software \fw.\\
    \ATR	&	Fixed memory buffer from which \authenticate reads \atoken and also where \install outputs \aack \\
    \IVTR	&	Reserved memory region for the MCU's IVT \\
    \SR	&	Fixed memory region where \status flag is stored; \status is used by \casusw for consistency. \\
    \bottomrule
    \vspace*{-3em}
    \end{tabularx}
\end{table}
    
Table \ref{table:notations} summarizes MCU hardware signals and memory regions relevant to \casu.
Figure \ref{fig:arch} illustrates the \casu architecture: (1) \casuhw prevents modification of memory regions in gray and prevents execution of all other memory, while (2) \casusw resides in the ROM; it contains a bootloader and subroutines related to secure update.
We describe these features in detail in the rest of this section.

\subsection{\casuhw: Hardware Security Monitor}
\casuhw monitors \pc, \wen, \daddr, \dmaen, \dmaaddr to detect illegal writes or execution.
When a \textit{violation} is detected, \casuhw activates the \reset signal.
To simplify notation when describing \casuhw properties, we define the following macro:

\vspace{-0.3cm}
{\footnotesize
\begin{equation*}
\modmem(i) \equiv (\wen \land \daddr = i) \lor (\dmaen \land \dmaaddr = i)
\end{equation*}
}
\vspace{-0.5cm}

$i$ represents a memory address. \noindent\modmem$(i)$ is true whenever the MCU core or the DMA is writing to $i$.
When representing a write within some contiguous memory region (with multiple addresses) $M = [M_{min}, M_{max}]$, 
we ``abuse'' the notation as \modmem$(M)$. To denote that a write has occurred within one of the multiple contiguous memory regions,
e.g., when a write happens to some address within $M_1$ or $M_2$, we say \modmem$(M_1,M_2)$.

\subsubsection{\bf Authorized Software Immutability} 
%


Software authorized by \casu, including any ISRs, is located in the contiguous memory segment \ER.
The pointer \EP stores the boundaries that define $ER$, i.e., \ermin and \ermax. 
\casuhw monitors \EP to locate the currently authorized software and enforce its rules based on this region. 
Write attempts to \EP are also monitored and only allowed when performed by \casusw, preventing malicious changes to \EP that could misconfigure the definition \ER, leading \casuhw to enforce protections based on the incorrect region. 
\ER is configurable to give \casusw flexibility to change the location and size of authorized software, instead of fixing \fw to the same location and size of \ow, as software versions vary in size.
\casuhw also protects memory regions \SR and \IVTR. 
\SR is used during a secure update, described in Section~\ref{sec:update_details}. 
Since ISRs are a part of \ER, IVT must be protected to maintain the integrity of interrupt handling during authorized software execution. 

Incidentally, Authorized Software Immutability also prohibits self-modifying code, i.e., 
code in \ER writing to \ER, to prevent code injection attacks within \ER.

\begin{figure}
    \centering
    \small
    \captionsetup{justification=centering}
    \begin{mdframed}
        \bf \small Authorized Software Immutability:
        \begin{equation}\label{eq:hwprop_wp}
            \begin{split}
                [\modmem(\ER,\EP,\SR,\IVTR) \land (PC \notin \TCR)] \rightarrow \reset
            \end{split}
        \end{equation}
        \bf \small Unauthorized Software Execution Prevention:
        \begin{equation}\label{eq:hwprop_usep}
            \begin{split}
                [(PC \notin \ER) \land (PC \notin \TCR)] \rightarrow \reset
            \end{split}
        \end{equation}
    \end{mdframed}
    \vspace{-0.8em}
    \caption{\casuhw Security Properties. 
    }\label{fig:hwprop}
    \vspace{-0.8em}
    \normalsize
\end{figure}

\subsubsection{\bf Unauthorized Software Execution Prevention}
Only authorized software (located in \ER) or \casusw (located in \TCR) are allowed to execute on \prv. 
Since \ER is configurable via \EP, after a secure update, \casusw re-configures \EP to allow execution from the new \ER location.

\subsubsection{\bf \casuhw Properties Formally}
Figure \ref{fig:hwprop} formalizes the aforementioned \casuhw security properties using proportional logic. 
Note that these properties must hold at all times. Equation \ref{eq:hwprop_wp} states that any modification to \ER, \EP, \SR, and \IVTR -- when a program other than \casusw ($PC \notin \TCR$) is executing --  causes a \reset. 
Equation \ref{eq:hwprop_usep} states that MCU cannot execute programs other than those in \ER and \TCR. 
If \pc points to any other memory location, the MCU is reset.

\subsection{\casu Secure Update}~\label{sec:update_details}

\noindent Recall (from Section \ref{subsec:su_scheme}) that \casu Secure Update  implements: (\update, \verify) on \vrf and (\authenticate, \install) on \prv.
At a high level, there are two ways of implementing it on \prv. 
\begin{compactenum}
\item Download \fw to DMEM (RAM), i.e., the stack or heap of the current software (\ow), and invoke \authenticate. 
If it succeeds, \install overwrites \ER with \fw and updates \EP.  
This is problematic, because, if a reset occurs in the middle of \install execution, then \ER containing \ow would be partially overwritten and \fw in the DMEM would be lost as a consequence of the reset. 
This would leave \prv software in a corrupted state.
\item Download \fw to PMEM (flash) and invoke \authenticate. 
If \authenticate succeeds, \install updates \EP to the location where \fw resides. 
This is generally safer since \fw and \ow reside in two separate flash memory regions. 
If the installation is interrupted by a reset, \casusw can re-invoke \install to complete the installation. 
However, this requires \prv PMEM to be sufficiently large to accommodate both \fw and \ow, i.e., at least double the size of \ER. 
We believe that this is a realistic assumption. The size of flash memory on our targetted devices is at least $8KB$, whereas 
the typical binary size is usually under $2KB$. 
\end{compactenum}
\noindent Construction \ref{construction:secure_update} shows the whole scheme.
Recall that \casusw is immutable (being in ROM). Its functionality is described below.

\begin{figure*}
    \captionsetup{justification=centering}
    \begin{mdframed}
        \begin{construction}\label{construction:secure_update}
            \small
            \casu Secure Update scheme defined by $[\update, \authenticate, \install, \verify]$ is realized as follows:
            ~\\~
                -- \key is a symmetric key pre-shared between \vrf and \prv (protected by \vrased secure architecture);
            \begin{compactenum}
            
                \item{$\update(\fw) \rightarrow \atoken$:\\}
                \vrf generates a tuple $T: =(\fw, \atoken)$, where \fw is the new software and \atoken 
                is the accompanying authentication token, as follows: 
                \begin{compactenum}
                    \item Compiles and generates \fw := $(L_{\fw}||V_{\fw}||N_{\fw}||BIN_{\fw}||IVT_{\fw})$, where $L_{\fw}$ 
                    is \fw size, $V_{\fw}$ is \fw version number, $N_{\fw}$ is a random nonce, $BIN_{\fw}$ 
                    is \fw binary, and $IVT_{\fw}$ is \fw \IVT,  to be placed in \IVTR of \prv.
                    \item Computes \atoken using equation \ref{eq:atoken} with the second operand set to: $0||\fw$, 
                    where '0' is the direction indicator from \vrf to \prv.
                    \begin{equation}\label{eq:atoken}
                        \atoken := HMAC(\key, 0||\fw)
                    \end{equation} 
                \end{compactenum}
                \vrf sends $T$ to \prv for update.
            ~
                \item{$\authenticate(\fw,\atoken) \rightarrow \perp/\top$:} \\
                Upon receiving a tuple $T: =(\fw, \atoken)$ from \vrf, \fw is downloaded at memory region pointed 
                to by \bEP and \atoken is written to \ATR. Then \prv does the following:
                \begin{compactenum}
                    \item If $V_{\fw} <= V_{\ER}$, output $\perp$ and return to \ER; otherwise, proceed to the next step.
                    \item Computes $\sigma$ using equation \ref{eq:sigma}.
                    \begin{equation}\label{eq:sigma}
                        \sigma := HMAC(\key, 0||\bEP)
                    \end{equation} 
                    \item If $\sigma == \atoken$, output $\top$ and invoke \install; otherwise, output $\perp$ and return to \ER, 
                    where the current software (\ow) resides.
                \end{compactenum}
            ~  
                \item{$\install(\fw) \rightarrow \aack$:} \\
                Upon invocation by \authenticate, or at boot time, in case \status is equal to $1$, \prv does the following:
                \begin{compactenum}
                    \item Sets \status to $1$ and updates \EP with values in \bEP.
                    \item Updates \IVTR with $\IVT_{\fw}$.
                    \item Computes \aack using equation \ref{eq:aack} and stores it at \ATR. In equation \ref{eq:aack} 
                    the second operand is $1||V_{\fw}||N_{\fw}$, where '1' is the direction indicator from \prv to \vrf.
                    \begin{equation}\label{eq:aack}
                        \aack := HMAC(\key, 1||V_{\fw}||N_{\fw})
                    \end{equation}
                    \item Sets \status to $0$ and jumps to new \ER, which is pointed to by the new value in \EP.
                \end{compactenum}
                \prv replies to \vrf with \aack indicating successful update.
            ~
                \item{$\verify(\aack) \rightarrow \perp/\top$:} \\
                Upon receiving \aack from \prv, \vrf does the following:
                \begin{compactenum}
                    \item Computes $\gamma$ using the same equation \ref{eq:aack}.
                    \item If $\gamma == \aack$, outputs $\top$; otherwise outputs $\perp$.
                \end{compactenum}
            \end{compactenum}
        \end{construction}
    \end{mdframed}
    \vspace{-0.8em}
    \normalsize
\end{figure*}

\subsubsection{\update}
Secure update requires for any software \fw to be installed on \prv to adhere to the following format 
$\fw := (L_{\fw}||V_{\fw}||N_{\fw}||BIN_{\fw}||IVT_{\fw})$, where $L_{\fw}$, $V_{\fw}$, $N_{\fw}$ is the \fw \textit{header} 
consisting of its size, version number, and a random nonce, respectively. 
$BIN_{\fw}$ is the \fw binary in byte-code that mandatorily includes a \download and \acknowledge subroutine 
that accepts future update requests and replies acknowledgment message back to \vrf. 
$IVT_{\fw}$ is the IVT of \fw that needs to be overwritten to \IVTR region so that MCU knows where to jump into the new software when an interrupt is triggered. 
Another requirement is that $V_{\fw}$ should always be greater than the version number of the current (or old) software on \prv. 
This avoids replay attacks that attempt to trick \prv into installing an old software version that contains vulnerabilities. 
In case \vrf wishes to revert to an older version (e.g., due to later-discovered bugs in \fw), it must issue a brand new update 
request with the older-version software, though with a {\bf new version} number.

\vrf, by invoking \update, computes \atoken using equation \ref{eq:atoken} and sends $(\fw, \atoken)$ to \prv. 

\subsubsection{\authenticate} \label{subsec:su_auth}
When \prv receives \update with \fw and \atoken, the current \download subroutine on \ow in \ER accepts and downloads \fw to 
an available PMEM slot. It then writes the pointers to \fw to \bEP, buffer Executable Pointer, in PMEM, and writes \atoken to \ATR. 
This \download subroutine should not be a part of \casusw, as exposing network interfaces directly to trusted parts of the device is hazardous and may result in the exploitation of unknown vulnerabilities in it, leading to key leakage. 
Hence, even though \ER is untrusted, it should be the one receiving the request, because even if it fails to receive or chooses to not call \authenticate, then \aack is not generated/sent, which is a clear indication to \vrf that the update was unsuccessful.

To securely verify that \fw is a valid software to be installed on \prv, \authenticate first checks whether the $V_{\fw}$ is greater than the one of \ER, i.e., $V_{\ER}$. 
If the $V_{\fw}$ is valid, it invokes \vrased as a subroutine to compute $\sigma$ according to equation \ref{eq:sigma}.  
If $\sigma$ matches with \atoken received from \vrf, then it outputs $\top$ (accept symbol) and further invokes \install to apply the update. 
Otherwise, it outputs $\perp$ (reject symbol) and returns to old software at \ER without computing any response to be sent back to \vrf. 

Note that \casusw execution is guarded by \casuhw (which inherits \vrased hardware properties), i.e., any interrupts or DMA, or any attempts to access the key or any confidential data that \casusw generates, will be considered as a \textit{violation} and an MCU reset will be triggered immediately. 
Also note that if such an abrupt reset occurs, MCU will return to the old software, and eventually \vrf has to send a new update request. 
In this new request, \vrf can use the same version number (but with a different nonce for maintaining freshness) because the previous update was not applied, and thus, the version number of the current software is still old.

\subsubsection{\install}
Once \fw is authenticated, \install is invoked.
This is the critical step of Secure Update. 
It is responsible for updating the \EP with \bEP, \IVTR with $\IVT_{\fw}$ and computing authenticated acknowledgment \aack that is to be replied to \vrf. 
As mentioned in Section \ref{subsec:su_auth}, if a reset occurs during any of these sub-steps, they have to be repeated from the beginning. 
This is because, if \EP is updated and \IVTR is not, vulnerabilities in old ISRs pointed to by the old \IVT can be exploited by malware. 
Furthermore, if \EP and \IVTR are updated, yet the computation of \aack failed, \vrf assumes that the update failed and repeats the update request with the same version number (since \EP is updated to the new software), and \authenticate will fail again. 
Therefore, all three sub-steps must take place atomically. 
To this end, \casusw uses a \status flag \SR in PMEM, which it sets and unsets, before and after the completion of \install sub-steps, respectively.

To handle cases when a reset is triggered during \install, the Reset Vector in \IVTR is programmed to start executing from \casusw. 
This technique is analogous to having a bootloader. At boot time, \casusw uses \status to determine whether a reset occurred prior to the completion of \install. 
If so, \casusw re-invokes \install from the beginning. 

Finally, \install computes \aack according to equation \ref{eq:aack} and writes it to \ATR.
After generating \aack, \casusw jumps to new \ER.  Now, it is the responsibility of the \acknowledge subroutine in \fw to reply to \vrf with \aack.

\noindent {\bf Acknowledgment Receipt:} There are two unlikely cases where \vrf may not receive \aack, after being generated by \install. 
Firstly, \aack sent by \prv being lost or corrupted in transit. 
In this case, upon a time-out, \vrf re-sends \update. 
Since \install stores \aack in a dedicated region of DMEM (\ATR), \download in \ER checks whether the update request has the same version number as itself and directly replies \aack to \vrf, instead of invoking \authenticate again. 
Secondly, a reset occurring after a successful update and before \aack is sent to \vrf.
In that case, \aack is lost and, upon a timeout, \vrf needs to send a new \update with a new version number. 
The drawback of this approach is that the same update is re-applied, wasting MCU clock cycles.
However, the latter case is very rare, and even if it occurs, \casusw only takes less than a second to re-install \fw (see Section \ref{subsec:runtime}). 

\vrf can distinguish between these cases by first re-sending the same \update.
If there is still no response, then \aack is most likely lost due to a reset and \vrf must send a new \update with a new version number. 




There are other ways to mitigate the aforementioned \aack issues. 
Rather than storing \aack in DMEM, it could be placed into a reserved memory in PMEM to ensure its persistence even if a reset occurs. 
Now, \download can always reply with \aack whenever it sees a duplicate request, thus eliminating the cost of re-update. 
However, this approach requires an additional write to flash, which may be undesirable. 
Alternatively, we can use a \vrf-supplied timestamp instead of a nonce in \fw and modify \authenticate 
to accept duplicate requests with a more recent timestamp. 
This approach does not require any reserved memory (not even in DMEM). 
However, it incurs runtime overhead every time \vrf issues a duplicate request.
Each aforementioned alternative has its own benefits and drawbacks. 
We leave it up to \vrf to decide which is most suitable. 

Note that none of the above can result in a DoS attack due to multiple requests,
because all \update-s originate from a legit \vrf and are verified by \authenticate. 
Moreover, \download can check the \fw header to check if the request was already seen,  
discard the rest of the packets, and simply reply stored \aack to \vrf.

\subsubsection{\verify}
Finally, if all goes well, \vrf receives an \aack and checks its validity verifies using equation \ref{eq:aack}. 
If either \aack is invalid, or a time-out occurs, \vrf assumes that the update failed.

\begin{figure}
	\centering
    \includegraphics[width=0.85\columnwidth]{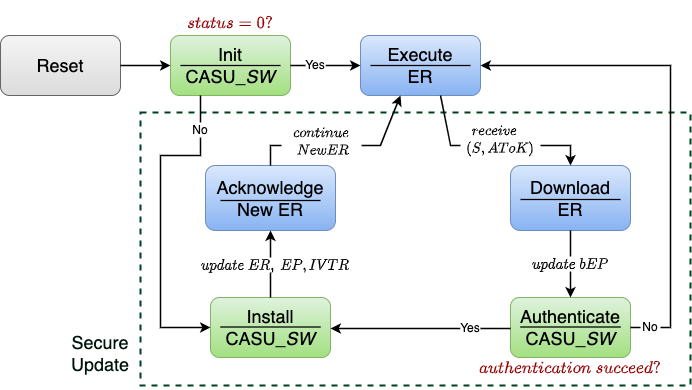}
    \caption{Secue Update Workflow: blue and green boxes indicate authorized and trusted execution routines, respectively. 
        }
    \label{fig:workflow}
\end{figure}

Figure \ref{fig:workflow} depicts the workflow of secure updates. When \prv comes out of reset, it starts executing \casusw. 
\casusw first checks whether \status is 1, it invokes \install to resume installation of already verified \fw located at \bEP. 
Otherwise, it jumps to \ow in \ER. Upon receiving \update, the \download routine in \ow accepts and downloads 
\fw to an available memory slot in PMEM and stores this address in \bEP.  \ow is free to complete its pending tasks 
before invoking \authenticate in \casusw.  Once, it invokes \casusw, atomic execution of \authenticate and \install (if the former 
succeeds) begins. During \install, if a \textit{violation} is detected, \prv resets and invokes \casusw with \status set to 1, thus
invoking \install again. After successful completion of \install, \casusw jumps to \fw in \ER. 
Eventually, the \acknowledge in \fw replies \aack to \vrf, and continues with its normal execution.

\section{Implementation}\label{sec:implementation}

\begin{figure}
	\centering
    \captionsetup{justification=centering}
    \includegraphics[width=0.9\columnwidth]{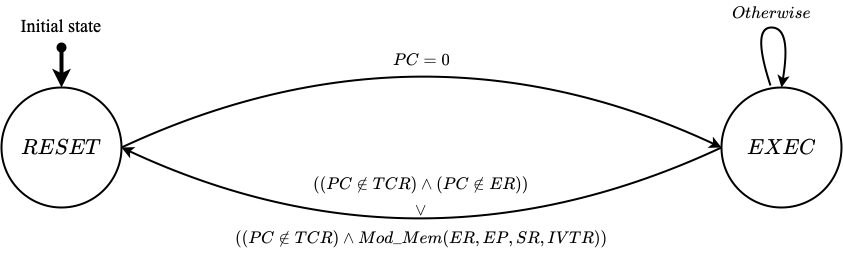}
    \caption{FSM of \casuhw Verified Hardware Module.}
    \label{fig:wp_fsm}
\end{figure}

\subsection{\casuhw Verified Hardware Module}
Figure \ref{fig:wp_fsm} presents a hardware FSM formally verified to enforce both properties of Figure \ref{fig:hwprop}. 
It is a Mealy FSM, where output is determined by both the current state and current input. 
This FSM takes as input the signals shown in Figure \ref{fig:arch} and produces a single one-bit output \reset. 
If \reset is $1$, the MCU core immediately resets.

There are two states in the FSM: \textit{RESET} and \textit{EXEC}. In \textit{RESET}, \reset is $1$ and remains so 
until the FSM leaves that state; in other cases reset is $0$. After a reset, as soon as \pc reaches $0$ (execution is ready to start), 
the FSM transitions to \textit{EXEC}. While in \textit{EXEC}, the FSM constantly checks for: {\it (1)}  modifications to \ER, \EP, \SR, or \IVTR,
and {\it (2)} execution attempts outside \ER and \TCR. In either case, the FSM transitions to \textit{RESET}.

We implement the FSM using Verilog HDL and automatically translate it into Symbolic Model Verifier (SMV) language using 
Verilog2SMV \cite{irfan2016verilog2smv} tool. Finally, we use the NuSMV Model Checker \cite{nusmv} to generate 
machine proofs showing that the FSM adheres to the properties in Figure \ref{fig:hwprop}. 

\subsection{\casusw Secure Update Routine}
\casusw implements subroutines \texttt{casu\_entry}, \texttt{casu\_authenticate}, \texttt{casu\_install}, and \texttt{casu\_exit}.

\texttt{casu\_entry} is the only legal entry point to \casusw; it is invoked at boot and during an update. 
Boot invocation is obtained by setting the IVT reset vector to \texttt{casu\_entry}.
\texttt{casu\_entry} takes a boolean argument to test whether it was invoked at boot or by \ER for an update. 
In the former case, it checks \status to determine whether to invoke \texttt{casu\_install} in order to resume the 
unfinished update from the last reset. Otherwise, it calls \texttt{casu\_exit}, which clears the MCU registers and 
jumps to the binary in \ER. 
In the latter case, it invokes \texttt{casu\_authenticate}.
\texttt{casu\_authenticate} checks for the validity of the version number of \fw at \bEP and invokes \vrased 
software to compute HMAC. If the measurement matches \atoken, \texttt{casu\_install} is invoked; 
otherwise, it jumps to \texttt{casu\_exit}. Finally, \texttt{casu\_install} updates \EP, copies the new IVT to \IVTR, and 
computes and stores \aack at \ATR. It also sets/unsets \status to indicate the status of installation to 
\texttt{casu\_entry} subroutine, in case of a reset.

\casusw is implemented in C with a tiny TCB of $\approx140$ lines of code. 
It uses \vrased software, which is implemented using a formally verified cryptographic library, HACL* \cite{hacl}. 

\section{Evaluation}\label{sec:evaluation}
All \casu source code and hardware verification/proofs are publicly available at \cite{casu-repo}.
\casu prototype is built on OpenMSP430 \cite{OpenMSP430}, an open-source implementation of TI-MSP430 \cite{TI-MSP430}. 
We use Xilinx Vivado to synthesize an RTL description of \casuhw and deploy it on the Diligent Basys3 board featuring an Artix7 FPGA.

\subsection{Hardware Overhead}
Table~\ref{tab:overhead_results} presents \casu hardware overhead compared to unmodified OpenMSP430 and \vrased. 
Similar to prior work~\cite{vrasedp,apex,sancus,smart}, we consider additional Look-Up Tables (LUTs) and registers.
Compared to \vrased, \casu only requires 3\% (99) additional LUTs and 0.3\% (34) additional registers.

\noindent {\bf Verification Cost:} \casu was verified using a Ubuntu 18.04 LTS machine running 3.2GHz with 16GB of RAM. 
Table~\ref{tab:overhead_results} shows verification time and memory. 
\casu requires $95$ additional lines of Verilog code to enforce properties in Figure \ref{fig:hwprop}. 
The verification cost includes the verification of \vrased properties. The time to verify the composite design is under
a second and requires 148MB of RAM.

\begin{table}[!htp]
\footnotesize
\centering
\caption{Hardware Overhead \& Verification cost.}
	\resizebox{\columnwidth}{!}{ 
		\begin{tabular}{|l|cc|cccc|} \hline\cline{1-7}
			\multirow{2}{*}{Architecture} & \multicolumn{2}{c|}{Hardware} & \multicolumn{4}{c|}{Verification}      \\
			                      & LUTs     & Regs     & LoC     & \#(LTLs)  &Time~(s)   & RAM~(MB)  \\ \hline
			OpenMSP430            & 1859     & 692      & -       &   -       & -         & -         \\
			\vrased               & 1902     & 724      & 481     &  10       & 0.4       &  13.6     \\
		    \casu(+\vrased)       & 1958     & 726      & 576     &  12       & 0.9       &  148       \\
			\hline\cline{1-7}
		\end{tabular}%
	}
	\label{tab:overhead_results}
\end{table}

\begin{figure}
	\centering
	\subfigure[Additional HW overhead (\%) in Number of Look-Up Tables]
	{\includegraphics[width=0.49\columnwidth]{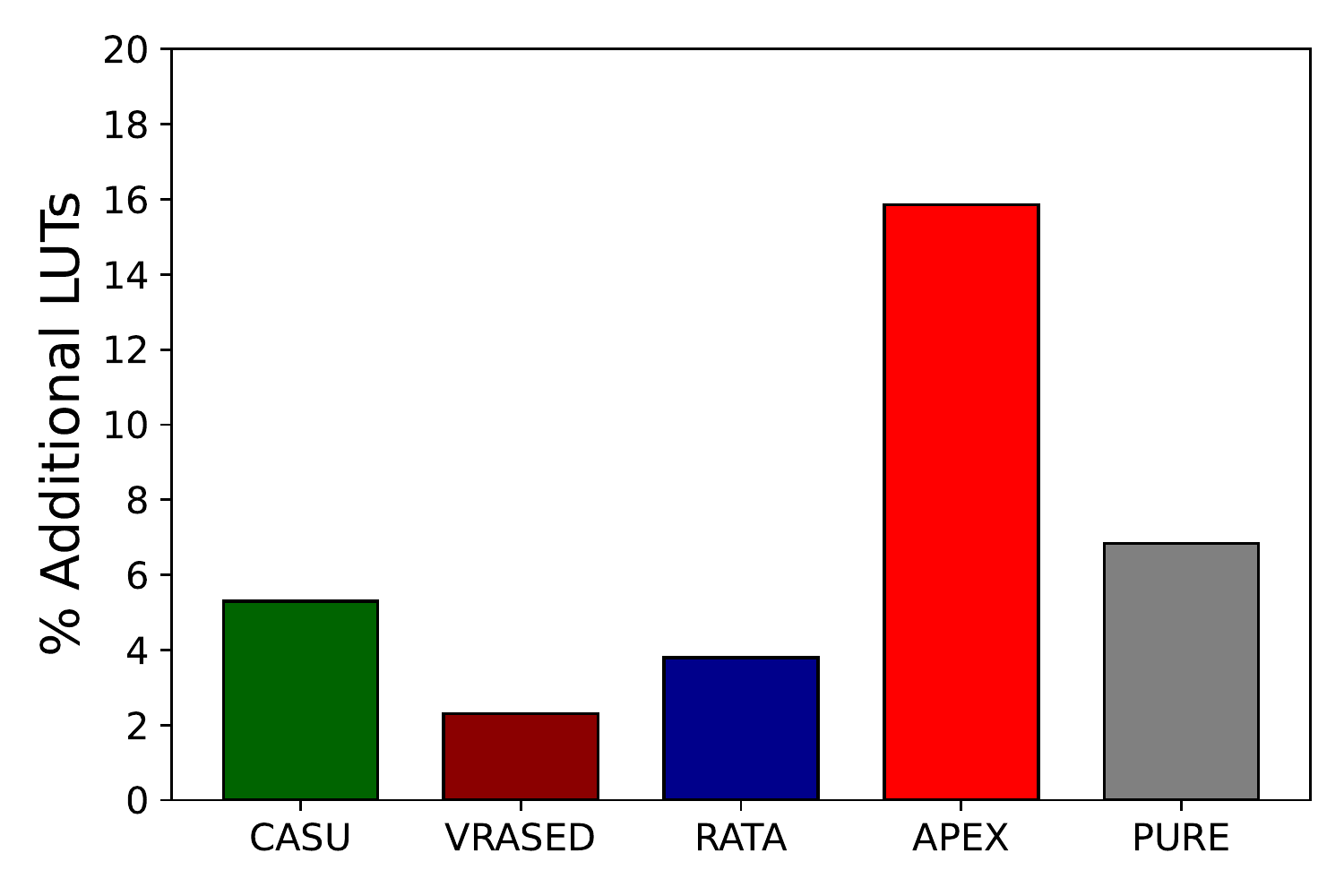}}
	\subfigure[Additional HW overhead (\%) in Number of Registers]
	{\includegraphics[width=0.49\columnwidth]{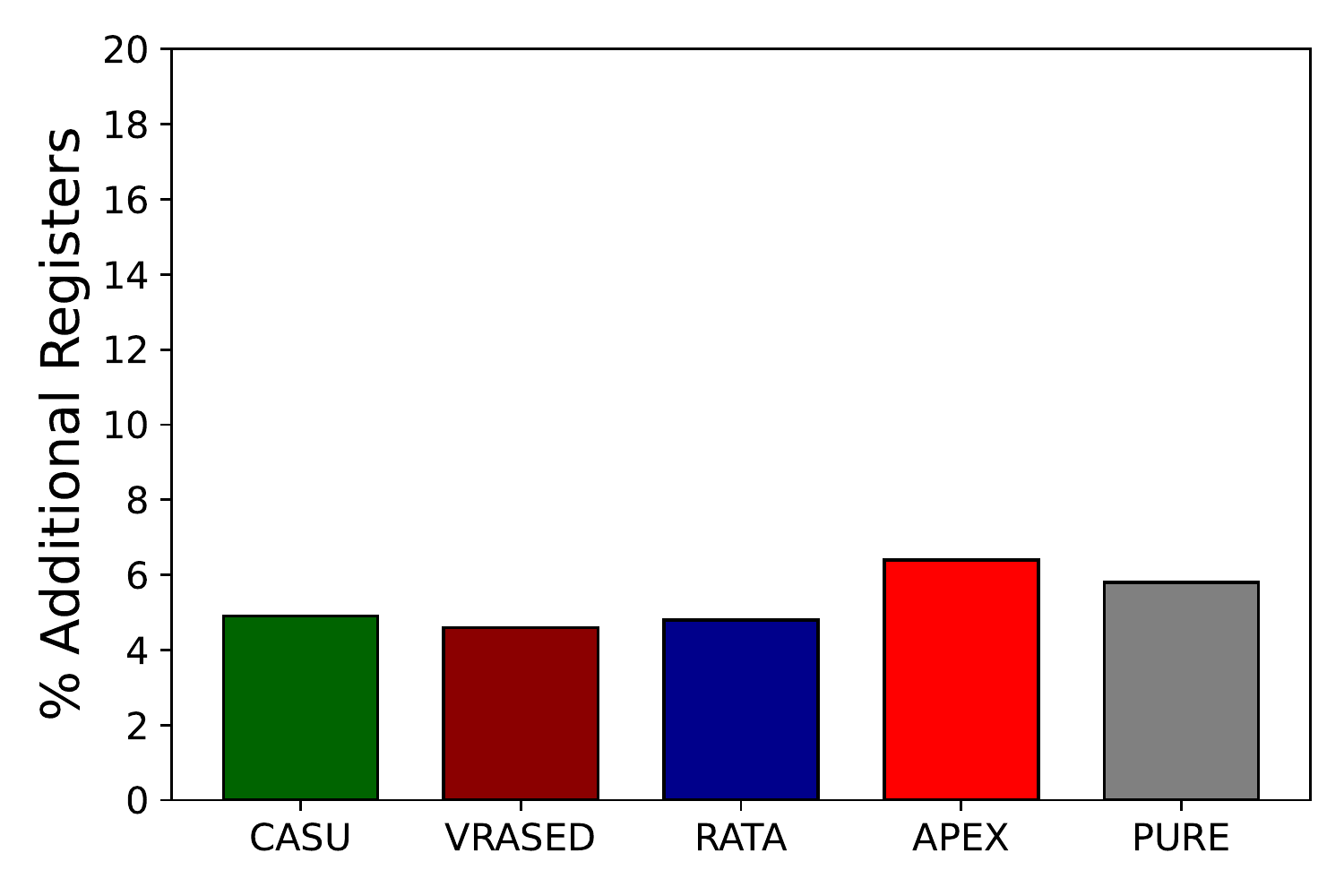}}
	\caption{Hardware Overhead Comparison.}
	\label{fig:hw_comparison} 
\end{figure}

\noindent {\bf Comparison with Related Architectures:} 
In Figure \ref{fig:hw_comparison}, we compare \casu with other low-end MCU security architectures, including 
\vrased~\cite{vrasedp}, \rata~\cite{rata}, \apex~\cite{apex}, and \pure~\cite{pure}, which provide \RA-related 
services. However, recall that, unlike \casu, all these other architectures are reactive. As a superset of \vrased, 
\casu naturally has a higher overhead. 
\casu and \rata have similar overheads, since both monitor memory modifications.
Whereas \apex and \pure enforce additional hardware properties for generating proofs of execution (\apex), and proofs of update, reset, erasure (\pure); and thus, they have a higher overhead than \casu. 

\subsection{Runtime for Secure Updates} \label{subsec:runtime}
\begin{figure}
	\centering
	\includegraphics[width=0.67\columnwidth]{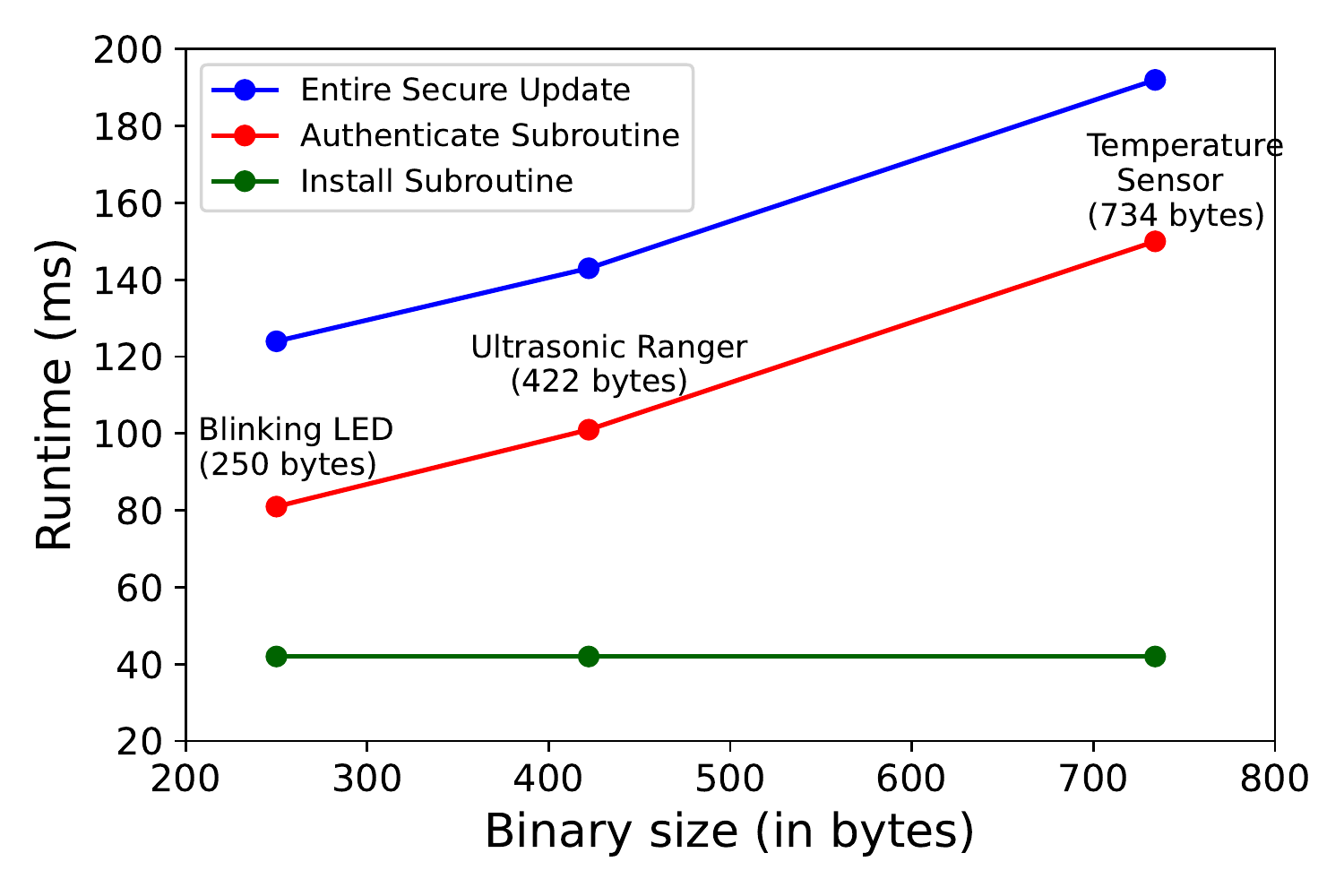}
	\caption{Runtime of \casusw Secure Update} \label{fig:runtime}
\end{figure}

The runtime of \casusw was evaluated on three sample applications: 
(1) Blinking LED (250 bytes of binary size) - toggles an LED every half a second, (2) Ultrasonic Ranger (422 bytes) - available at \cite{ultrasonic_ranger} - computes the distance of an obstacle from a moving object, and (3) Temperature Sensor (734 bytes) - available at \cite{temp_humi_sensor} - measures the temperature of a room. 
In each case, we measured execution time of \texttt{casu\_authenticate} and \texttt{casu\_install} -- the most time-consuming tasks dominated by \hmac computations.
Results are shown in Figure \ref{fig:runtime}. 
\texttt{casu\_install} runtime is constant because it updates fixed-size memory ranges (including \EP, \IVTR, and \SR) and computes \hmac on a fixed-size input.
Whereas, \texttt{casu\_authenticate} scales linearly with \fw size, over which \hmac is computed.
The combined runtime for the worst case (temperature sensor case with 734-byte binary) is $\approx 200ms$, which we consider to be reasonable, considering that updates are infrequent.

\noindent {\bf Reserved Memory:} 
\casu requires 32 bytes of reserved RAM for \ATR, 8 bytes of reserved PMEM for \EP and \bEP, and 1 byte of PMEM for \SR. 
In total, it consumes 41 bytes of additional storage.

\section{Related Work}\label{sec:relatedwork}
Prior related work generally falls into two categories: \textit{passive} and \textit{active} Roots-of-Trust (RoTs). 

\noindent {\bf Passive RoTs} 
aim to detect software compromise by producing an unforgeable proof of \prv state to \vrf.
In terms of functionality, they implement the following services: (1) memory integrity verification, i.e., 
\RA~\cite{smart,sancus,vrasedp,simple,tytan,trustlite,TPM,KeJa03,seshadri2004swatt,SLS+05,SLP08,GGR09,flicker,SCHELLEKENS200813}; 
(2) verification of runtime properties, including control-flow and data-flow attestation~\cite{apex,litehax,cflat,lofat,atrium,oat,tinycfa,dialed,geden2019hardware};
and (3) proofs of remote software update, erasure, and reset~\cite{pure,verify_and_revive,asokan2018assured}.
As mentioned in Section \ref{sec:intro}, they are passive in nature and do not prevent modifications. 
Whereas, \casu is active and, as such, ensures software immutability except for authorized updates. 
However, \casu is similar to these \RA techniques with respect to updates. 

\noindent {\bf Active RoTs} proactively monitor \prv behavior to prevent (or minimize the extent of) compromises.
For example, \cite{awdt-dominance,lazarus-effect,garota} are architectures that guarantee execution of critical tasks 
even when all other software is compromised. Similarly, VERSA~\cite{pfb} guarantees sensor data privacy for low-end MCUs by allowing only authorized software to access and process sensed quantities. In contrast, \casu can be viewed as an active RoT that focuses on software 
immutability, prevention of illegal execution, and authorized updates.

\noindent {\bf Remote Over-the-Air (OTA) Updates} 
support seamless delivery of software updates for IoT devices. Notably, TUF~\cite{tuf} is an update delivery framework 
resilient to key compromises. Uptane~\cite{uptane} extends TUF for supporting updates for vehicular ECUs. 
However, both TUF and Uptane require relatively heavy cryptographic operations, unsuitable for \casu-targeted low-end devices. 
ASSURED \cite{asokan2018assured} extends TUF to provide a secure update framework for large-scale IoT deployments. 
SCUBA \cite{scuba} uses software-based attestation to identify and patch infected software regions. However, due to the timing assumptions of software-based attestation, it is unsuitable for remote IoT settings. 
PoSE~\cite{perito2010secure} and AONT~\cite{secu} use proofs of secure erasure to wipe \prv to show that its
memory is fully erased and then install new software. However, these schemes are not fault-tolerant and can not retain previous software, in case of reset during erasure or new update installation.  
Also, an extensive discussion of various software update schemes can be found in~\cite{update-survey}.

\noindent {\bf Formal Verification} 
provides increased confidence about the correctness of security techniques' implementations. 
In the space of low-end MCUs, \vrased~\cite{vrasedp} and \rata~\cite{rata} are formally verified hybrid \RA architectures, where the latter one detects \toctou attacks.
APEX \cite{apex} and PURE \cite{pure} offer formally verified proofs of remote software execution, and proof of update, reset, and erasure. 
Similarly, \casu offers a verified hardware module for authorized software immutability and unauthorized execution prevention.

\vspace{-1pt}
\section{Conclusions}\label{sec:conclusion}
In this paper, we designed \casu, a prevention-based root-of-trust architecture for low-end MCUs. 
\casu differs from prior work by disallowing illegal software modifications rather than detecting them. 
\casu also prevents execution of any unauthorized software and supports secure software updates. 
\casu is prototyped on OpenMSP430 and its hardware component is 
formally verified. 
Experiments show that \casu incurs quite low overhead and is thus suitable for resource-constrained low-end IoT devices. 
Its entire implementation is publicly available at \cite{casu-repo}.

\balance
\bibliographystyle{ieeetr}
\bibliography{references}

\begin{thebibliography}{10}

\bibitem{FC08}
A.~Francillon and C.~Castellucia, ``Code injection attacks on
  harvard-architecture devices,'' in {\em CCS '08}, 2008.

\bibitem{runtime_attacks_sok}
L.~Szekeres, M.~Payer, T.~Wei, and D.~Song, ``Sok: Eternal war in memory,'' in
  {\em 2013 IEEE Symposium on Security and Privacy}, pp.~48--62, IEEE, 2013.

\bibitem{cowan2000buffer}
C.~Cowan, F.~Wagle, C.~Pu, S.~Beattie, and J.~Walpole, ``Buffer overflows:
  Attacks and defenses for the vulnerability of the decade,'' in {\em IEEE
  DISCEX}, IEEE, 2000.

\bibitem{owasp_attacks}
OWASP, ``Owasp top ten.'' \url{https://owasp.org/www-project-top-ten/}, 2021.

\bibitem{smart}
K.~Eldefrawy, G.~Tsudik, A.~Francillon, and D.~Perito, ``{SMART}: Secure and
  minimal architecture for (establishing dynamic) root of trust,'' in {\em
  NDSS}, 2012.

\bibitem{sancus}
J.~Noorman, J.~V. Bulck, J.~T. M{\"{u}}hlberg, F.~Piessens, P.~Maene,
  B.~Preneel, I.~Verbauwhede, J.~G{\"{o}}tzfried, T.~M{\"{u}}ller, and F.~C.
  Freiling, ``Sancus 2.0: {A} low-cost security architecture for iot devices,''
  {\em {ACM} Trans. Priv. Secur.}, vol.~20, no.~3, pp.~7:1--7:33, 2017.

\bibitem{vrasedp}
I.~De~Oliveira~Nunes, K.~Eldefrawy, N.~Rattanavipanon, M.~Steiner, and
  G.~Tsudik, ``{VRASED}: A verified hardware/software co-design for remote
  attestation,'' in {\em USENIX Security}, 2019.

\bibitem{simple}
M.~Ammar, B.~Crispo, and G.~Tsudik, ``Simple: A remote attestation approach for
  resource-constrained iot devices,'' in {\em 2020 ACM/IEEE 11th International
  Conference on Cyber-Physical Systems (ICCPS)}, pp.~247--258, IEEE, 2020.

\bibitem{tytan}
F.~Brasser, B.~E. Mahjoub, A.~Sadeghi, C.~Wachsmann, and P.~Koeberl, ``Tytan:
  tiny trust anchor for tiny devices,'' in {\em Proceedings of the 52nd Annual
  Design Automation Conference, San Francisco, CA, USA, June 7-11, 2015},
  pp.~34:1--34:6, {ACM}, 2015.

\bibitem{trustlite}
P.~Koeberl, S.~Schulz, A.-R. Sadeghi, and V.~Varadharajan, ``{TrustLite}: A
  security architecture for tiny embedded devices,'' in {\em EuroSys}, 2014.

\bibitem{pure}
I.~De~Oliveira~Nunes, K.~Eldefrawy, N.~Rattanavipanon, and G.~Tsudik, ``Pure:
  Using verified remote attestation to obtain proofs of update, reset and
  erasure in low-end embedded systems,'' 2019.

\bibitem{verify_and_revive}
M.~Ammar and B.~Crispo, ``Verify\&revive: Secure detection and recovery of
  compromised low-end embedded devices,'' in {\em Annual Computer Security
  Applications Conference}, pp.~717--732, 2020.

\bibitem{asokan2018assured}
N.~Asokan, T.~Nyman, N.~Rattanavipanon, A.-R. Sadeghi, and G.~Tsudik,
  ``{ASSURED}: Architecture for secure software update of realistic embedded
  devices,'' {\em IEEE Transactions on Computer-Aided Design of Integrated
  Circuits and Systems}, vol.~37, no.~11, 2018.

\bibitem{rata}
I.~De~Oliveira~Nunes, S.~Jakkamsetti, N.~Rattanavipanon, and G.~Tsudik, ``On
  the toctou problem in remote attestation,'' {\em CCS}, 2021.

\bibitem{atmel_specs}
``Avr atmega 1284p 8-bit microcontroller.''
  \url{http://ww1.microchip.com/downloads/en/DeviceDoc/doc8059.pdf}, 2009.

\bibitem{TI-MSP430}
T.~Instruments, ``Msp430 ultra-low-power sensing \& measurement mcus.''
  \url{http://www.ti.com/microcontrollers/msp430-ultra-low-power-mcus/overview.html}.

\bibitem{trustzone}
{Arm Ltd.}, ``Arm {TrustZone}.''
  \url{https://www.arm.com/products/security-on-arm/trustzone}, 2018.

\bibitem{sgx}
Intel, ``{I}ntel {Software Guard Extensions} ({Intel} {SGX}).''
  \url{https://software.intel.com/en-us/sgx}.

\bibitem{OpenMSP430}
O.~Girard, ``open{MSP430},'' 2009.

\bibitem{hacl}
J.-K. Zinzindohou{\'e}, K.~Bhargavan, J.~Protzenko, and B.~Beurdouche, ``Hacl*:
  A verified modern cryptographic library,'' in {\em CCS}, 2017.

\bibitem{DolevYao}
D.~Dolev and A.~Yao, ``On the security of public key protocols,'' {\em IEEE
  Transactions on Information Theory}, 1983.

\bibitem{ravi2004tamper}
S.~Ravi, A.~Raghunathan, and S.~Chakradhar, ``Tamper resistance mechanisms for
  secure embedded systems,'' in {\em VLSI Design}, 2004.

\bibitem{irfan2016verilog2smv}
A.~Irfan, A.~Cimatti, A.~Griggio, M.~Roveri, and R.~Sebastiani,
  ``{Verilog2SMV}: A tool for word-level verification,'' in {\em Design,
  Automation \& Test in Europe Conference \& Exhibition (DATE), 2016}, 2016.

\bibitem{nusmv}
A.~Cimatti, E.~Clarke, E.~Giunchiglia, F.~Giunchiglia, M.~Pistore, M.~Roveri,
  R.~Sebastiani, and A.~Tacchella, ``Nusmv 2: An opensource tool for symbolic
  model checking,'' in {\em CAV}, 2002.

\bibitem{casu-repo}
``{CASU} source code.'' \url{https://github.com/sprout-uci/CASU}, 2022.

\bibitem{apex}
I.~De~Oliveira~Nunes, K.~Eldefrawy, N.~Rattanavipanon, and G.~Tsudik, ``{APEX}:
  A verified architecture for proofs of execution on remote devices under full
  software compromise,'' in {\em 29th {USENIX} Security Symposium ({USENIX}
  Security 20)}, (Boston, MA), {USENIX} Association, Aug. 2020.

\bibitem{ultrasonic_ranger}
``Ultrasonic ranger code.''
  \url{https://github.com/Seeed-Studio/LaunchPad_Kit/tree/master/Grove_Modules/ultrasonic_ranger}.

\bibitem{temp_humi_sensor}
``Temperature sensor code.''
  \url{https://github.com/Seeed-Studio/LaunchPad_Kit/tree/master/Grove_Modules/temp_humi_sensor}.

\bibitem{TPM}
{Trusted Computing Group.}, ``Trusted platform module (tpm),'' 2017.

\bibitem{KeJa03}
R.~Kennell and L.~H. Jamieson, ``Establishing the genuinity of remote computer
  systems,'' in {\em USENIX Security Symposium}, 2003.

\bibitem{seshadri2004swatt}
A.~Seshadri, A.~Perrig, L.~Van~Doorn, and P.~Khosla, ``{SWATT}: Software-based
  attestation for embedded devices,'' in {\em IEEE Symposium on Research in
  Security and Privacy (S\&P)}, (Oakland, California, USA), pp.~272--282, IEEE,
  2004.

\bibitem{SLS+05}
A.~Seshadri, M.~Luk, E.~Shi, A.~Perrig, L.~van Doorn, and P.~Khosla, ``Pioneer:
  {V}erifying code integrity and enforcing untampered code execution on legacy
  systems,'' in {\em ACM SOSP}, 2005.

\bibitem{SLP08}
A.~Seshadri, M.~Luk, and A.~Perrig, ``{SAKE}: {S}oftware attestation for key
  establishment in sensor networks,'' in {\em DCOSS}, 2008.

\bibitem{GGR09}
R.~W. Gardner, S.~Garera, and A.~D. Rubin, ``Detecting code alteration by
  creating a temporary memory bottleneck,'' {\em IEEE TIFS}, 2009.

\bibitem{flicker}
J.~M. McCune, B.~J. Parno, A.~Perrig, M.~K. Reiter, and H.~Isozaki, ``Flicker:
  An execution infrastructure for tcb minimization,'' in {\em Proceedings of
  the 3rd ACM SIGOPS/EuroSys European Conference on Computer Systems 2008},
  pp.~315--328, 2008.

\bibitem{SCHELLEKENS200813}
D.~Schellekens, B.~Wyseur, and B.~Preneel, ``Remote attestation on legacy
  operating systems with trusted platform modules,'' {\em Science of Computer
  Programming}, vol.~74, no.~1, pp.~13 -- 22, 2008.

\bibitem{litehax}
G.~Dessouky, T.~Abera, A.~Ibrahim, and A.-R. Sadeghi, ``Litehax: lightweight
  hardware-assisted attestation of program execution,'' in {\em 2018 IEEE/ACM
  International Conference on Computer-Aided Design (ICCAD)}, pp.~1--8, IEEE,
  2018.

\bibitem{cflat}
T.~Abera, N.~Asokan, L.~Davi, J.~Ekberg, T.~Nyman, A.~Paverd, A.~Sadeghi, and
  G.~Tsudik, ``{C-FLAT:} control-flow attestation for embedded systems
  software,'' in {\em Proceedings of the 2016 {ACM} {SIGSAC} Conference on
  Computer and Communications Security, Vienna, Austria, October 24-28, 2016}
  (E.~R. Weippl, S.~Katzenbeisser, C.~Kruegel, A.~C. Myers, and S.~Halevi,
  eds.), pp.~743--754, {ACM}, 2016.

\bibitem{lofat}
G.~Dessouky, S.~Zeitouni, T.~Nyman, A.~Paverd, L.~Davi, P.~Koeberl, N.~Asokan,
  and A.-R. Sadeghi, ``Lo-fat: Low-overhead control flow attestation in
  hardware,'' in {\em Proceedings of the 54th Annual Design Automation
  Conference 2017}, p.~24, ACM, 2017.

\bibitem{atrium}
S.~Zeitouni, G.~Dessouky, O.~Arias, D.~Sullivan, A.~Ibrahim, Y.~Jin, and A.-R.
  Sadeghi, ``Atrium: Runtime attestation resilient under memory attacks,'' in
  {\em Proceedings of the 36th International Conference on Computer-Aided
  Design}, pp.~384--391, IEEE Press, 2017.

\bibitem{oat}
Z.~Sun, B.~Feng, L.~Lu, and S.~Jha, ``Oat: Attesting operation integrity of
  embedded devices,'' in {\em 2020 IEEE Symposium on Security and Privacy
  (SP)}, pp.~1433--1449, IEEE, 2020.

\bibitem{tinycfa}
I.~De~Oliveria~Nunes, S.~Jakkamsetti, and G.~Tsudik, ``{Tiny-CFA}: Minimalistic
  control-flow attestation using verified proofs of execution,'' in {\em
  Design, Automation and Test in Europe Conference (DATE)}, 2021.

\bibitem{dialed}
I.~De~Oliveira~Nunes, S.~Jakkamsetti, and G.~Tsudik, ``Dialed: Data integrity
  attestation for low-end embedded devices,'' 2021.

\bibitem{geden2019hardware}
M.~Geden and K.~Rasmussen, ``Hardware-assisted remote runtime attestation for
  critical embedded systems,'' in {\em 2019 17th International Conference on
  Privacy, Security and Trust (PST)}, pp.~1--10, IEEE, 2019.

\bibitem{awdt-dominance}
M.~Xu, M.~Huber, Z.~Sun, P.~England, M.~Peinado, S.~Lee, A.~Marochko,
  D.~Mattoon, R.~Spiger, and S.~Thom, ``Dominance as a new trusted computing
  primitive for the internet of things,'' in {\em 2019 {IEEE} Symposium on
  Security and Privacy, {SP} 2019, San Francisco, CA, USA, May 19-23, 2019},
  pp.~1415--1430, {IEEE}, 2019.

\bibitem{lazarus-effect}
M.~Huber, S.~Hristozov, S.~Ott, V.~Sarafov, and M.~Peinado, ``The lazarus
  effect: Healing compromised devices in the internet of small things,'' in
  {\em {ASIA} {CCS} '20: The 15th {ACM} Asia Conference on Computer and
  Communications Security, Taipei, Taiwan, October 5-9, 2020} (H.~Sun,
  S.~Shieh, G.~Gu, and G.~Ateniese, eds.), pp.~6--19, {ACM}, 2020.

\bibitem{garota}
E.~Aliaj, I.~De~Oliveira~Nunes, and G.~Tsudik, ``{GAROTA:} generalized active
  root-of-trust architecture,'' {\em CoRR}, vol.~abs/2102.07014, 2021.

\bibitem{pfb}
I.~De~Oliveira~Nunes, S.~Hwang, S.~Jakkamsetti, and G.~Tsudik,
  ``Privacy-from-birth: Protecting sensed data from malicious sensors with
  {VERSA},'' {\em CoRR}, vol.~abs/2205.02963, 2022.

\bibitem{tuf}
J.~Samuel, N.~Mathewson, J.~Cappos, and R.~Dingledine, ``Survivable key
  compromise in software update systems,'' in {\em Proceedings of the 17th ACM
  Conference on Computer and Communications Security}, p.~61–72, Association
  for Computing Machinery, 2010.

\bibitem{uptane}
T.~Karthik, A.~Brown, S.~Awwad, D.~McCoy, R.~Bielawski, C.~Mott, S.~Lauzon,
  A.~Weimerskirch, and J.~Cappos, ``Uptane: Securing software updates for
  automobiles,'' in {\em International Conference on Embedded Security in Car},
  pp.~1--11, 2016.

\bibitem{scuba}
A.~Seshadri, M.~Luk, A.~Perrig, L.~van Doorn, and P.~Khosla, ``Scuba: Secure
  code update by attestation in sensor networks,'' in {\em In Proceedings of
  the 5th ACM workshop on Wireless security (WiSe '06)}, p.~85–94, 2006.

\bibitem{perito2010secure}
D.~Perito and G.~Tsudik, ``Secure code update for embedded devices via proofs
  of secure erasure.,'' in {\em ESORICS}, 2010.

\bibitem{secu}
G.~O. Karame and W.~Li, ``Secure erasure and code update in legacy sensors,''
  in {\em Trust and Trustworthy Computing}, pp.~283--299, Springer
  International Publishing, 2015.

\bibitem{update-survey}
K.~Zandberg, K.~Schleiser, F.~Acosta, H.~Tschofenig, and E.~Baccelli, ``Secure
  firmware updates for constrained iot devices using open standards: A reality
  check,'' {\em IEEE Access}, pp.~71907--71920, 2019.

\end{thebibliography}
\end{document}